\documentclass{jfp1}

\newcommand\bcmdtab{\noindent\bgroup\tabcolsep=0pt%
  \begin{tabular}{@{}p{10pc}@{}p{20pc}@{}}}
\newcommand\ecmdtab{\end{tabular}\egroup}

\usepackage{listings}
\usepackage{verbatim}
\usepackage{todonotes}
\usepackage{inconsolata}
\usepackage{mathtools}
\usepackage[htt]{hyphenat}
\usepackage[inline,shortlabels]{enumitem}
\usepackage{hyperref}

\title[Journal of Functional Programming]
      {parboiled2: a macro-based approach for effective generators of parsing expressions grammars in Scala}

 \author[Alexander A. Myltsev]
        {\textbf{Alexander A. Myltsev}\\
         Moscow Institute of Physics and Technology\\
         Faculty of Applied Mathematics and Computer Science\\
         Department of Discrete Mathematics\\
         141701 Dolgoprudny, Russia\\
         e-mail: alexander [dot] myltsev [at] phystech.edu}

\jdate{October 2016}
\pubyear{2016}
\pagerange{\pageref{firstpage}--\pageref{lastpage}}
\doi{???}

\newcommand{\emphcode}[1]{\texttt{#1}}

\lstdefinelanguage{scala}{
  morekeywords={%
          abstract,case,catch,class,def,do,else,extends,%
          false,final,finally,for,forSome,if,implicit,import,lazy,%
          match,new,null,object,override,package,private,protected,%
          return,sealed,super,this,throw,trait,true,try,type,%
          val,var,while,with,yield},
  otherkeywords={=>,<-,<\%,<:,>:,\#,@},
  sensitive=true,
  morecomment=[l]{//},
  morecomment=[n]{/*}{*/},
  morestring=[b]",
  morestring=[b]',
  morestring=[b]""",
  basicstyle=\ttfamily
}[keywords,comments,strings]
\begin{document}

\label{firstpage}

\maketitle

\begin{abstract}

In today's computerized world, parsing is ubiquitous. Developers parse logs,
queries to databases and websites, programming and natural languages. When Java
ecosystem maturity, concise syntax, and runtime speed matters, developers choose
\emph{parboiled2} that generates grammars for parsing expression grammars
(PEG). The following open source libraries have chosen \emph{parboiled2}
for parsing facilities:

\begin{itemize}
\item \emph{akka-http} is the Streaming-first HTTP server/module of Lightbend Akka
\item \emph{Sangria} is a Scala GraphQL implementation
\item \emph{http4s} is a minimal, idiomatic Scala interface for HTTP
\item \emph{cornichon} is Scala DSL for testing HTTP JSON API
\item \emph{scala-uri} is a simple Scala library for building and parsing URIs
\end{itemize}

The library uses a wide range of Scala facilities to provide required
functionality. We also discuss the extensions to PEGs. In particular, we show
the implementation of an internal Scala DSL that features intuitive syntax and
semantics. We demonstrate how \emph{parboiled2} extensively uses Scala typing
to verify DSL integrity. We also show the connections to inner structures of
\emph{parboiled2}, which can give the developer a better understanding of how
to compose more effective grammars. Finally, we expose how a grammar is expanded
with Scala Macros to an effective runtime code.

\end{abstract}

% \tableofcontents

\section{Introduction}

Computer specialists have been parsing programming languages and protocols
since the beginning of the computer era. They used Noam Chomsky's generative system
of grammars, context-free grammars (CFGs), and regular expressions (REs) to
encode syntax of programming languages and protocols. One of the purposes of
generative grammars was to model natural languages and hence to inherit ambiguity
in their design. The uncertainty of CFGs brings unnecessary complexity to parsing in
machine languages that are explicit by design. There are several alternatives
to CFGs to specify syntax formally.

Parser combinators \cite{wadler1995monads, moors2008parser} are popular due to
their readability, modularity, and ease of maintenance, they
cannot be fully used in production. The first reason is that naive
implementations do not handle left-recursive grammars, unless they are
implemented according to a solution given in
\cite{Frost:2007:MET:1621410.1621425}. Another reason lies in
the expressive power that causes runtime inefficiency because of the
composition overhead and the creation of intermediate data structures. 
A significant performance speedup is given in \cite{beguet2014accelerating}
by removing overheads and deleting intermediate data
representations. The authors used meta-programming techniques such as
macros~\cite{Burmako:2013:SML:2489837.2489840} and
staging~\cite{rompf2010lightweight}.

Parsing Expression Grammars (PEGs) are an alternative solution to the parsing
problem. The difference with CFGs is that PEGs eliminate the ambiguity by
\emph{prioritized choice} in the process of recognition-based syntax describing
\cite{Ford:2004:PEG:982962.964011}. Virtually, PEGs make a suitable
replacement for REs \cite{Mozzherin2017}. PEGs work as fast as REs-based
parsers (even faster in some edge cases). The benefit is that PEGs allow
natural parsing of sequences that are defined recursively (XML, JSON,
programming languages, etc.). Finally, PEGs are much easier to read and
maintain than REs, recognize left
recursion~\cite{Medeiros:2014:LRP:2844735.2844770}, support
backtracking~\cite{Redziejowski:2007:PEG:1366071.1366090}, and semantic
actions~\cite{Atkey:2012:SPS:2358958.2359493}.

In the paper, we describe the implementation of the \emph{parboiled2} library. The
\emph{parboiled2} is an implementation of PEGs parsers generators in the Scala
programming language \cite{Odersky:2016:PSU:2988396}. \emph{parboiled2} is
assembled as a regular Java Virtual Machine (JVM) library. Any JVM-oriented
development environment, profiler, debugger, tracer, etc. can use the library.

The paper
\begin{enumerate*}[label={\alph*)}]
  \item develops intuition about how to use PEGs with the \emph{parboiled2} DSL,
  \item exposes inner structure of the library,
  \item explains tight connections of the library inner parts,
  \item describes how macro generates a fast runtime code,
  \item lists current limitations of the library.
\end{enumerate*}

Section~\ref{section-inner-implementation} of the paper describes the core
parts of a simple \emph{parboiled2} grammar.  Section~\ref{rules-dsl}
introduces a high-level domain specific language (DSL) to describe rules of
recognition. Section~\ref{semantics-of-parsing} provides insight into the
parsing process and describes its semantics in detail.
Section~\ref{parsing-actions-and-value-stack} explains how \emph{parboiled2}
produces side effects with \emph{Value Stack}. Section~\ref{typing-of-rule-dsl}
explains how \emph{parboiled2} uses a Scala type checker to verify every rule and
their composition. Section~\ref{code-generation} explains the process of
step-by-step code generation of macro definitions
\cite{Burmako:2013:SML:2489837.2489840}. Section~\ref{catching-parsing-errors}
exposes the way how \emph{parboiled2} catches and handles parsing errors.

\section{Implementation of Inner Abstractions}\label{section-inner-implementation}

A parser for a particular grammar should be derived from the \emphcode{Parser} base
class to inherit all the necessary facilities to parse input string. The
\emphcode{Parser} inheritor expects the \emphcode{input} of type
\emphcode{ParserInput} in the constructor. \emph{parboiled2} provides implicit
conversions from three types to \emphcode{ParserInput} type: \emphcode{String}
by default, \emphcode{Array[Char]}, and \emphcode{Array[Byte]}.

Consider a PEG that recognizes mathematical formulas of four basics operations
with precedence to non-negative integers (Fig.~\ref{peg-calculator}). The
corresponding \emph{parboiled2} parser is shown in
Fig.~\ref{parboiled2-peg-calculator}. \emphcode{CalculatorParser} is a Scala
class. It contains a composition of rules that determine the parsing process.
Note that \emphcode{Expression} (Fig.~\ref{parboiled2-peg-calculator}, line 3)
has an explicit type since it is recursively used in \emphcode{Factor}
(Fig.~\ref{parboiled2-peg-calculator}, line 9). All the rules bodies start with
a \emphcode{rule} method call. The call body contains a composition of built-in
rules from the DSL and calls to other \emphcode{CalculatorParser} rules in the
scope.

\begin{figure}
\[
\begin{array}{lcl}
expression  & \leftarrow & term~((\verb"'+'"~/~\verb"'-'")~term)* \\
term        & \leftarrow & factor~((\verb"'*'"~/~\verb"'/'")~factor)* \\
factor      & \leftarrow & number~/~(\verb"'('"~expression~\verb"')'") \\
number      & \leftarrow & [0-9]+
\end{array}
\]
\unprogrammath
\caption{PEG for mathematical formulas of four operations to non-negative
integers}\label{peg-calculator}
\end{figure}

\begin{figure*}
\centering
\begin{lstlisting}[language=Scala, numbers=left, escapechar=\%]
class CalculatorParser(val input: ParserInput) extends Parser {
  def InputLine = rule { Expression ~ EOI }
  def Expression: Rule1[Expr] = rule {
    Term ~ ('+' ~ Term %\underline{$\sim> Add$}% | '-' ~ Term %\underline{$\sim> Sub$}%).*
  }
  def Term = rule {
    Factor ~ ('*' ~ Factor %\underline{$\sim> Mul$}% | '/' ~ Factor %\underline{$\sim> Div$}%).*
  }
  def Factor = rule { Number | '(' ~ Expression ~ ')' }
  def Number = rule { %\underline{$capture$}%(CharPredicate.Digit.+) %\underline{$\sim> Val$}% }
}
\end{lstlisting}
\caption{\emph{parboiled2} rules for mathematical formulas}\label{parboiled2-peg-calculator}
\end{figure*}

\begin{figure*}
\centering
\begin{lstlisting}[language=Scala]
sealed trait Expr
case class Val(value: String)        extends Expr
case class Add(lhs: Expr, rhs: Expr) extends Expr
case class Sub(lhs: Expr, rhs: Expr) extends Expr
case class Mul(lhs: Expr, rhs: Expr) extends Expr
case class Div(lhs: Expr, rhs: Expr) extends Expr
\end{lstlisting}
\caption{AST for mathematical formulas grammar}\label{parboiled2-calculator-ast}
\end{figure*}

The grammar in Fig.~\ref{peg-calculator} is only recognized if the input string
is an arithmetic expression. To be useful in practice, a parser performs
semantic actions such as computing an expression or emitting AST nodes. With the
underlined code parts on lines 4, 7, 10 (Fig.~\ref{parboiled2-peg-calculator})
\emphcode{CalculatorParser} captures the input parts and produces AST nodes
listed in Fig.~\ref{parboiled2-calculator-ast}.

\emphcode{CalculatorParser} successfully parses the input string, and the result
returned contains AST nodes in the Scala interpreter as follows:

\begin{lstlisting}
scala> new CalculatorParser("1+(2-3*4)/5").InputLine.run()
res0: scala.util.Try[Expr] =
  Success(Add(Val(1),Mul(Val(2),Val(3))))
\end{lstlisting}

If parsing fails, it returns the \emphcode{Failure} of \emphcode{ParseError} type.
\emphcode{ParseError} contains all the necessary information about errors to
print a comprehensive string message that describes why the parsing failed.
The following example shows \emphcode{ParseError} generation by feeding an invalid
string to a \emphcode{CalculatorParser} constructor:

\begin{lstlisting}
scala> val parser = new CalculatorParser("1+2!3")
scala> val Failure(e: ParseError) = parser.InputLine.run()
e: org.parboiled2.ParseError =
  ParseError(Position(3,1,4), Position(3,1,4), <6 traces>)
scala> parser.formatError(e)
res1: String =
Invalid input '!', expected '/', '+', '*', 'EOI', '-'
or Digit (line 1, column 4):
1+2!3
\end{lstlisting}

Since the PEGs are recognition-based, a parser should define a rule
\emphcode{Expression$\sim$EOI} that would force the parser to move to the end of
the input string. Otherwise, the parser successfully parses the arithmetic
expression \emph{"1+2"} until it encounters an unexpected char \emph{'!'}:

\begin{lstlisting}
scala> new CalculatorParser("1+2!3").Expression.run()
res2: scala.util.Try[Expr] = Success(Add(Val(1),Val(2)))
\end{lstlisting}

\section{Rules DSL}\label{rules-dsl}

\emphcode{CalculatorParser} contains a composition of elementary rules that are
listed in Table~\ref{parboiled2-rules}. The rules are naturally grouped into
three categories: \emphcode{basic}, \emphcode{combinators}, and \emphcode{semantic
actions}. These categories are defined in the corresponding Scala traits
\emphcode{RuleDSLBasics}, \emphcode{RuleDSLCombinators},
\emphcode{RuleDSLActions}. \emphcode{basic} and \emphcode{combinators} rules
are derived from the original definition of PEGs
\cite{Ford:2004:PEG:982962.964011}. \emphcode{semantic actions} allow a parser
to produce useful results (like the AST of the parsed expression).

\emph{parboiled2} directs a developer to program a statically correct
grammar with two facilities.

The first facility against usage errors of the library is the
\emphcode{rule} macro. We designed every \emph{parboiled2} rule call to exist
only within the \emphcode{rule} macro scope. If a rule is called somewhere outside
of the macro, the Scala compiler fails with an error. Practically every rule
has an annotation that prevents it from existing at compile-time. The
\emphcode{rule} macro erases the rule calls by expanding their composition to a
runnable code.

The second facility is the type system that helps to verify if a
\emphcode{rule} can be run against the input. For example,
\emphcode{Expression} has a type stating that it returns an AST node of type
\emphcode{Expr}. Hence, the entire rule composition of \emphcode{Expression}
body should be of type \emphcode{Expr}.
Section~\ref{parsing-actions-and-value-stack} shows more sophisticated examples.

\begin{table*}
  \caption{parboiled2 rules}
  \label{parboiled2-rules}
  \begin{minipage}{\textwidth}
    \centering
    \setlength{\tabcolsep}{2pt}
    \begin{tabular}{lllr}
      Rule category & parboiled2 rule & PEG operator& Description \\
      \hline
      Basic rules   & ch\footnote{Extended by \emphcode{ignoreCase(c: Char)} that matches
                                   an input char case insensitively}
                             & ' '           & Literal character \\
                    & str\footnote{Extended by \emphcode{ignoreCase(s: String)} that matches
                                    an input string case insensitively} 
                             & " "           & Literal string \\
                    & CharPredicate\footnote{An efficient implementation composable
                              of character sets. It comes with a number pre-defined character
                              classes like \emphcode{CharPredicate.Digit} or \emphcode{CharPredicate.LowerHexLetter}}
                              & [ ] & Character class \\

                    & ANY\footnote{Generalized to \emphcode{anyOf(chars: String)} that matches any char of provided ones.
                                   \emphcode{noneOf(chars: String)} is an inversion of \emphcode{anyOf} -- fails on any of provided
                                   chars }& . & Any character \\
                    & & EOI & Matches end of line \\
                    & ( )                         & (e)           & Grouping \\
                    & anyOf()        & & \\
      Combinator rules & $e_1 \sim e_2$ & $e_1 \sim e_2$ & Sequence \\
                       & $e_1 | e_2$ & $e_1 | e_2$ & Prioritized Choice (First Of) \\
                       & optional(e)   & e?            &  Optional \\
                       & zeroOrMore(e) & e*            & Zero-or-more \\
                       & oneOrMore(e)  & e+            & One-or-more \\
                       & \&(e) & \&e & And-predicate \\
                       & !(e) & !e & Not-predicate \\
      Semantic actions & $\xrightarrow{\mathbf{f}_n}$ & & Action operator \\
                       & push(value)                  & & Pushes the value to the ValueStack \\
                       & drop                         & & Drops a value from the ValueStack \\
                       & capture(e)                   & & Pushes captured string to the ValueStack \\
    \end{tabular}
    % \vspace{\baselineskip}
  \end{minipage}
\end{table*}

\section{Semantics of Parsing} \label{semantics-of-parsing}

PEG parsers are recursive-descent parsers with backtracking. Most parsers
produced by traditional parser generators like ANTLR have two parsing phases,
whereas PEGs have only one parsing phase. PEGs do not require any look-ahead, and
they perform quite well in most real-world scenarios. However, certain pathological
languages implemented in PEGs and inputs exhibit exponential runtime
\cite{Ford:2004:PEG:982962.964011}.

When the runner executes a rule against the current position in an input
buffer, the rule applies its specific matching logic to the input. When a
\emphcode{Parser} calls a \emphcode{rule} method, it creates an instance of
a \emphcode{ParserState} class that stores reference to the input and the
\emphcode{cursor} of the \emphcode{Int} type. The \emphcode{cursor} points to the
next unmatched input character. In case of successful parsing by rule, the
parser advances the cursor and potentially executes the next rule.
Otherwise, when the rule fails, the cursor is reset to the last successful
match. And the parser backtracks in search for another parsing alternative that
might succeed.

Consider this simple \emph{parboiled2} rule:

\begin{lstlisting}
def foo = rule { 'a' ~ ('b' ~ 'c' | 'b' ~ 'd') }
\end{lstlisting}

When the rule attempts to match against the input \verb'"abd"', the parser performs
the following steps:

\begin{enumerate}[\#1.]
  \item Rule \emphcode{foo} starts executing, which calls its first sub-rule
    \emphcode{'a'}. The cursor sets to position~0.

  \item Rule \emphcode{'a'} is executed against the input at position~0,
    matches (succeeds), and the cursor advances to position~1.

  \item Rule \emphcode{'b'$\sim$'c'$\vert$'b'$\sim$'d'} starts executing, which
    calls its first sub-rule \emphcode{'b'$\sim$'c'}.

  \item Rule \emphcode{'b'$\sim$'c'} starts executing, which calls its first
    sub-rule \emphcode{'b'}.

  \item Rule \emphcode{'b'} is executed against the input position~1, matches
    (succeeds), and the cursor advances to position~2.

  \item Rule \emphcode{'c'} is executed against the input position~2 and mismatches (fails).

  \item Rule \emphcode{'b'$\sim$'c'$\vert$'b'$\sim$'d'} notices that its first
    sub-rule has failed, resets the cursor to position~1 and calls its second
    sub-rule \emphcode{'b'$\sim$'d'}.

  \item Rule \emphcode{'b'$\sim$'d'} starts executing, which calls its first
    sub-rule \emphcode{'b'}.

  \item Rule \emphcode{'b'} is executed against the input position~1, matches, and
    the cursor advances to position~2.

  \item Rule \emphcode{'d'} is executed against the input position~2, matches, and
    the cursor advances to position~3.

  \item Rule \emphcode{'b'$\sim$'d'} completes successfully, as its last
    sub-rule has succeeded.

  \item Rule \emphcode{'b'$\sim$'c'$\vert$'b'$\sim$'d'} completes successfully,
    as one of its sub-rules has succeeded.

  \item Rule \emphcode{foo} completes execution successfully, as its last
    sub-rule has succeeded. The whole input \emphcode{"abd"} is matched, and
    the cursor is left at position~3 (after the last-matched character).

\end{enumerate}

\section{Parsing Actions and Value Stack} \label{parsing-actions-and-value-stack}

The primary difference between \emph{parboiled2} and Scala combinator parsers
lies in the way they produce the result of parsing. Every Scala combinator
parsers grammar is a composition of functions: they always produce a result
that is then passed as an argument to another parsing function. The problem is
that parsing produces plenty of intermediate and mostly redundant data
structures that cause extra calls of memory allocations and garbage collections
in JVM. \cite{LiHaoyi2014} solved this problem by writing
an effective runtime Scala code. \cite{Jonnalagedda2014} used
compile-time staging to eliminate redundant data structures.
\emph{parboiled2} introduces a particular data structure called
\emphcode{ValueStack}: the library user should decide how to manipulate
intermediate parsing structures.

The \emphcode{ValueStack} is a mutable extension of \emphcode{Iterable[Any]} that
implements an untyped stack of values. \emphcode{Parser} creates a fresh new
instance of the \emphcode{ValueStack} upon every start rule run. It is a
private member of \emphcode{ParserState} of \emphcode{Parser}'s internals and
it is not intended to be used directly. We extend seven inductive definitions 
of \textit{parsing expressions} given in Section 3.1 of
\cite{Ford:2004:PEG:982962.964011} with \emphcode{semantic actions}
to operate on the \emphcode{ValueStack} in Fig.~\ref{semantic-actions-extension}.
In addition, we extend relation $\Rightarrow_{G}$
(Fig.~\ref{g-relation-extension}): from triples of the form $(e,x,S)$ to
triples of the form $(n,o,S')$, where $e,x,n,o$ are defined in
\cite{Ford:2004:PEG:982962.964011}. $S$ indicates the state of the
\emphcode{ValueStack} before a matching attempt. $S'$ is the state after a
matching attempt.

\begin{figure*}

\begin{enumerate}[1.]
  \item \emphcode{push} accepts one or more arguments and immediately
    pushes to the \emphcode{ValueStack}

  \item \emphcode{capture} accepts a rule as a single argument. If the
    provided rule succeeds to match, then the captured part of the input is
    pushed to the \emphcode{ValueStack}.

  \item $\xrightarrow{\mathbf{f}_n}$ (``action expression'') of arity $n$,
    pops values $v_1, v_2, \ldots, v_n$ ($v_n$ is assigned to the first popped
    value, $v_{n-1}$~-- to the next popped, etc.) from the \emphcode{ValueStack},
    applies a given function $\mathbf{f}_n(v_1, v_2, \ldots, v_n)$, and pushes
    the result back to the \emphcode{ValueStack}.

    Note the inversive order of sequential pops. We made this design decision
    to unify the usage of the \emphcode{ValueStack}. We suggest to keep in mind
    this memo: the \emphcode{ValueStack} grows from left to right, and arguments of
    the function $\mathbf{f}_n$ are assigned from left to right from the most
    recently pushed values that are at the right of the \emphcode{ValueStack}.

  \item \emphcode{zero-or-more(e,$\xrightarrow{\mathbf{f}_2}$)} can be
    defined using binary ``action expression'' that becomes a ``reduction
    expression''. While \emphcode{e} matches the input, the reduction
    \emphcode{zero-or-more} pops one value from the \emphcode{ValueStack} after
    every successful match of \emphcode{e}, applies $\mathbf{f}_2$ to it, and
    puts the result back to the \emphcode{ValueStack}.

\end{enumerate}

\caption{Extension to PEG: Semantic actions}
\label{semantic-actions-extension}

\end{figure*}

\begin{figure}

\begin{enumerate}
  \item \textbf{Standard expression}: $(e,x,S) \Rightarrow (n,o,S)$. Any
    standard expression does not change the state of the \emphcode{ValueStack}.

  \item \textbf{Push}: $(push(v),x,S) \Rightarrow (1,o,S.push(v))$

  \item \textbf{Capture (success case)}: If $(e,x,S) \Rightarrow (n,o,S)$, then
    $(capture(e),x,S) \Rightarrow (n+1,o,S.push(o))$

  \item \textbf{Capture (failure case)}: If $(e,x,S) \Rightarrow (n,f,S)$, then
    $(capture(e),x,S) \Rightarrow (n+1,f,S)$

  \item \textbf{Action expression (success case)}: If $(e,x,S) \Rightarrow
    (n,o,S)$, then $n$ values are popped from the \emphcode{ValueStack}: $v_n =
    S.pop(), v_{n-1}=S.pop(), \ldots v_1=S.pop()$ and $(e
    \leadsto_{\mathbf{f}},x,S) \Rightarrow (n+1,o,S.push(\mathbf{f}(v_1, v_2,
    \ldots v_n))$

  \item \textbf{Action expression (failure case)}: If $(e,x,S) \Rightarrow
    (n,f,S)$, then $(e \leadsto_{\mathbf{f}},x,S) \Rightarrow (n+1,f,S)$

  \item \textbf{Zero-or-more reduction (repetition case)} If $(e, x_1 x_2 y, S)
    \Rightarrow (n_1, x_1, S.push(x_1))$ and
    $(e^{*(\xrightarrow{\mathbf{f}_2})}, x_1 x_2 y, S) \Rightarrow (n_2, x_1
    x_2, S.push(x_1).push(x_2))$, then $(e^{*(\xrightarrow{\mathbf{f}_2})}, x_1
    x_2 y, S) \Rightarrow (n_1+n_2+1, x_1 x_2, S.push(\mathbf{f}_2(S.pop(),
    x_1)).push(\mathbf{f}_2(S.pop(), x_2)))$

  \item \textbf{Zero-or-more reduction (termination case)} if $(e, x, S)
    \Rightarrow (n_1, f, S)$, then $(e^{*(\xrightarrow{\mathbf{f}_2})},x,S)
    \Rightarrow (n_1+1, \varepsilon, S)$ 

\end{enumerate}

\caption{Extension to relation $\Rightarrow_{G}$}
\label{g-relation-extension}

\end{figure}

\section{Typing of Rule DSL} \label{typing-of-rule-dsl}

\emph{parboiled2} uses the Scala type system to catch potential problems at the
compile time in two ways as follows:

\begin{enumerate}
  \item \emph{parboiled2} start rule can return a result of one of three
    types based on imported \emphcode{DeliveryScheme} implicits.

  \item \emph{parboiled2} verifies access to the \emphcode{ValueStack} based on
    types of rules. It allows avoiding most of the inconsistent states of
    the \emphcode{ValueStack}.

\end{enumerate}

\subsection{DeliveryScheme of Parsing Result} \label{delivery_scheme-parsing_result}

\emphcode{run} method launches the start rule of a \emphcode{Parser} against
the provided input string. The parsing then could end in one of three possible
ways:

\begin{itemize}
  \item \emphcode{success} if the parser successfully matches the input. In this
    case, the parsing result should hold an instance subclass of shapeless
    \emphcode{HList}

  \item \emphcode{parseError} if the parser fails to match against the given input. In
    this case, parsing should return \emphcode{parboiled2.ParserError} that
    contains information on why the parsing failed

  \item \emphcode{error} if the parser fails for an internal reason (division by
    zero, index out of range, etc.). In this case, the parsing returns an instance
    of \emphcode{scala.Throwable} subclass

\end{itemize}

\emph{parboiled2} supports three ways to deliver the \emphcode{success/failure}
result: \emphcode{scala.util.Try}, \emphcode{scala.Either}, and simply throwing
an exception. We abstract it to \emphcode{Result} (embedded in
\emphcode{DeliveryScheme}) that has three instances -- one per type of the
result. Fig.~\ref{delivery-scheme} shows the implementation of
\emphcode{DeliveryScheme} for the \emphcode{scala.util.Try} result type.

The \emphcode{run} method implicitly accepts the particular instance of
\emphcode{DeliveryScheme} available in the scope of calling. It then internally
wraps the success or failure result by calling \emphcode{scheme} instance methods.

\begin{figure*}
  \centering
  \begin{lstlisting}[language=Scala]
trait DeliveryScheme[L <: HList] {
  type Result
  def success(result: L): Result
  def parseError(error: ParseError): Result
  def failure(error: Throwable): Result
}

implicit def Try[L <: HList] = new DeliveryScheme[L] {
  type Result = Try[L]
  def success(result: L) = Success(result)
  def parseError(error: ParseError) = Failure(error)
  def failure(error: Throwable) = Failure(error)
}

def run()(implicit scheme: Parser.DeliveryScheme[L]): scheme.Result = {
  val result: HList = // ...
  scheme.success(result)
}
  \end{lstlisting}
  \caption{Varying result type of \emphcode{run} configurable by implicitly provided deliver scheme}
  \label{delivery-scheme}
\end{figure*}

\subsection{Rule Types} \label{sect-rule-types}

The parsing process changes the \emphcode{ValueStack} as a side effect. Naive parsing
can lead the \emphcode{ValueStack} to an inconsistent state. For example, a rule
might pop a value from an empty stack, or cast a popped value to a wrong
type. The Scala type system prevents many invalid operations at the
type-checking phase of compilation.

We attach extra type information to \emphcode{Rule} that keeps track on how it
intends to change the \emphcode{ValueStack}. \emphcode{Rule} is isomorphic to
Scala functions: it accepts the input of a particular type from the
\emphcode{ValueStack} values and produces an output of another type that pushes
to the \emphcode{ValueStack}. From this perspective, \emphcode{Rule} is defined in
the same way as a regular function:
\lstinline[language=Scala]{class Rule[-I <: HList, +O <: HList]}, where \emphcode{I} and
\emphcode{O} are types of the input and the output, respectively.
For example, parser rules of the type
\lstinline[language=Scala]{Rule[Int::String::HNil,}
\lstinline[language=Scala]{String::HNil]} are only allowed to pop from the \emphcode{ValueStack}
value of the \emphcode{Int} type, then of the \emphcode{String} type (note
the order: \emphcode{Int} is first), and push a value only of the \emphcode{String} type.

\emph{Basic rules} are not intended to change the \emphcode{ValueStack}
(Fig.~\ref{parboiled2-rules}). They have the
\lstinline[language=Scala]{type} \lstinline[language=Scala]{Rule0 = Rule[HNil, HNil]}.

\emph{Action rules} change the \emph{ValueStack} in a straightforward way.
\emphcode{capture} and \emphcode{push} can only push values to the
\emphcode{ValueStack}. The \emphcode{push} rule pushes a value of any type
unconditionally. The \emphcode{capture} rule expects that the provided inner rule
matches, and only then it pushes the matched string. It is the moment where
the type-level computation happens: both \emphcode{capture} and \emphcode{push}
either decrease \emphcode{I} if it is not \emphcode{HNil}, or append to the
output type \emphcode{O} of the parent rule.

The complimentary \emphcode{drop} rule unconditionally pops and throws away one
or more values from the \emphcode{ValueStack}. It either decreases the \emphcode{O}
type, if it's not \emphcode{HNil}, or extend the input type \emphcode{I} of the
parent rule.

Action operator $\xrightarrow{\mathbf{f}_n}$ can either decrease or prepend
an additional type to either \emphcode{I} or \emphcode{O}. It depends on the
relation of how many arguments it simultaneously intends to pop and push to
the \emphcode{ValueStack}.

More sophisticated type-level computations stand behind \emph{rule
combinators}. A \emphcode{Sequence} combinator should check whether type
\emphcode{$O_{LHS}$} of the left-hand-side (LHS) rule is compatible with
\emphcode{$I_{RHS}$} of right-hand-side (RHS) rule, i.e., check if the LHS rule
pushes the values of types that RHS expects to pop from the \emphcode{ValueStack}.
If \emphcode{$O_{LHS}$} and \emphcode{$I_{RHS}$} are of different sizes, the
\emphcode{sequence combinator} then checks either \emphcode{$I_{LHS}$} or
\emphcode{$O_{RHS}$}, if it can handle a ``larger'' rule.

A special case of \emph{rule combinators} is the so-called \emphcode{reduction rule}.
Consider this common scenario of reducing the input string to a single value on
the \emphcode{ValueStack}. The rule \emphcode{Factor} from
Fig.~\ref{parboiled2-peg-calculator} is extended to handle multiplication
operation:

\begin{lstlisting}[language=Scala]
(Factor: Rule1[Int]) ~
  zeroOrMore('*' ~ Factor ~> ((a: Int, b) => a * b))
\end{lstlisting}

\emphcode{zeroOrMore} hosts two operations inside:

\begin{enumerate}
  \item it matches \emphcode{Factor} expression that pushes the value of type
    \emphcode{Int} according to its type

  \item then precisely in the same iteration of \emphcode{zeroOrMore} it pops
    \textbf{two} values from the \emphcode{ValueStack}, and pushes those values
    multiplication back to the \emphcode{ValueStack}

\end{enumerate}

The type of the inner rule is \lstinline[language=Scala]{Rule[Int::Int::HNil, Int::HNil]}.
Since \lstinline[language=Scala]{Int::HNil} is nested to
\lstinline[language=Scala]{Int::Int::HNil}, type of \emphcode{zeroOrMore} is
computed to \lstinline[language=Scala]{Rule[Int::HNil, Int::HNil]}

In total, starting from the empty \emphcode{ValueStack} and intending to leave it
empty or push some values to it, a custom rules composition of the \emphcode{Rule}
types mutually fulfill constraints:

\begin{itemize}

  \item the parsing ends with no values on the \emphcode{ValueStack}, i.e., the grammar
    recognizes an input. Or parsing stops with one or more values on the
    \emphcode{ValueStack}

  \item a rule that pops values of some types from the \emphcode{ValueStack}
    provides handling function of the same types

  \item none of the rules attempts to pop a value if the \emphcode{ValueStack} is
    empty

\end{itemize}

It is worth mentioning that Scala erases all types information during
compilation. It means that there is no overhead of any sophisticated types-casts at
runtime.

Next, we will describe some rules in detail. To keep the length of the paper
reasonable, we do not cover type signatures of all basic rules. We describe
several simple rules to give some intuition on how to read the rest of the
rules.

\begin{figure*}

  \begin{lstlisting}[language=Scala]

def capture[I <: HList, O <: HList](r: Rule[I, O])
  (implicit p: Prepend[O, String :: HNil]): Rule[I, p.Out]

def push[T](value: T)(implicit h: HListable[T]): Rule[HNil, h.Out]

object HListable extends LowerPriorityHListable {
  implicit def fromUnit: HListable[Unit] { type Out = HNil } = `n/a`
  implicit def fromHList[T <: HList]: HListable[T] { type Out = T } = `n/a`
}
abstract class LowerPriorityHListable {
  implicit def fromAnyRef[T]: HListable[T] { type Out = T :: HNil } = `n/a`
}
  \end{lstlisting}
  \caption{Type signature of basic rules}
  \label{rules-type-signatures}

\end{figure*}

\subsubsection{capture}

\emphcode{capture} (Fig.~\ref{rules-type-signatures}) pushes the matched
string to the \emphcode{ValueStack}. The \emphcode{capture} takes any valid rule
\lstinline[language=Scala]{r: Rule[I,O]} as an argument. The \emphcode{capture}
resulting input type is the same as the \emphcode{r}'s input type. \emphcode{capture}
prepends \emphcode{String} to \emphcode{r}'s output type:
\lstinline[language=Scala]{O::String::HNil} in pseudonotation.
\emph{shapeless} higher kinded \emphcode{Prepend} type of argument types
\emphcode{O} and \emphcode{String::HNil} computes \emphcode{capture}'s output
type and put it to \emphcode{p.Out}.

\subsubsection{push}

\emphcode{push} (Fig.~\ref{rules-type-signatures}) does not depend on any
inner rule. It pushes a value of an arbitrary type \emphcode{T}. The
complication arises from the cases of what type \emphcode{T} might be:

\begin{itemize}

  \item in case of \emphcode{Unit} nothing is pushed. \emphcode{push} that
    attempts to handle value of the \emphcode{Unit} type is equivalent to calling
    \emphcode{run}

  \item \emphcode{T<:HList}. Then all the values of HList are pushed as
    individual elements

  \item a single value of any other type \emphcode{T} is pushed as is

\end{itemize}

This pattern match on the type level is implemented in
\emphcode{parboiled2.support.HListable} type as follows. \emph{parboiled2}
defines three implicits with appropriate \emphcode{Out} types for each case:
\emphcode{Unit}, \emphcode{T<:HList} and low-priority \emphcode{AnyRef}.
Depending on type \emphcode{T} and implied type \emphcode{HListable[T]},
corresponding implicit with \emphcode{Out} type would be given to value
\emphcode{h} of \emphcode{push}. Defining \emphcode{fromAnyRef} as
\emphcode{LowerPriorityHListable} prevents its being given as an implicit for
\emphcode{h} of any type.

\subsubsection{sequence}

\emphcode{sequence} matches when both left-hand-side (LHS) and right-hand-side
(RHS) rules are matched. It implies that the LHS and RHS rules on
the \emphcode{ValueStack} should be compatible on the type level. There are three
possible cases:

\begin{itemize}

  \item when both rules pop nothing from the \emphcode{ValueStack}. No matter what they push
    to it (even no values), the result would be a concatenation of $O_{LHS}$ and
    $O_{RHS}$. In types (using abbreviated \emphcode{HList} pseudonotation) it is encoded:
    \lstinline[language=Scala]{Rule[,A]~Rule[,B]=Rule[,A:B]}

  \item \lstinline[language=Scala]{Rule[A:B:C,D:E:F]~Rule[F,G:H]=Rule[A:B:C,D:E:G:H]}
    type describes the case when the LHS rule pushes enough values to be popped
    by the RHS rule, no matter what the LHS rule actually pops, and the RHS
    pushes. The result type should be as it pops the LHS values, and the
    rightmost values of the LHS rules that are equivalent to the popped values
    of the RHS rule wiped out and replaced by the pushed values of the RHS rule

  \item and the final case is when the RHS rule pops more values than the LHS
    has pushed. In this case, the result rule demands the missing values to pop
    and leaves the pushed values as they are. The encoded type is
    \lstinline[language=Scala]{Rule[A,B:C]~Rule[D:B:C,E:F]=Rule[D:A,E:F]}.

\end{itemize}

The type-level implementation of the algorithm is listed in
Fig.~\ref{sequence-type-level-algorithm}.

\begin{figure*}
\centering
\begin{lstlisting}[language=Scala, showstringspaces=false]
@tailrec def rec(L, LI, T, TI, R, RI) =
  if (TI <: L) R
  else if (LI <: T) RI.reverse ::: R
  else if (LI <: HNil) rec(L, HNil, T, TI.tail, R, RI)
  else if (TI <: HNil) rec(L, LI.tail, T, HNil, R, LI.head :: RI)
  else rec(L, LI.tail, T, TI.tail, R, LI.head :: RI)
rec(L, L, T, T, R, HNil)
\end{lstlisting}
\caption{Type-level implementation of \emphcode{sequence} output type computation}
\label{sequence-type-level-algorithm}
\end{figure*}

\section{Code Generation} \label{code-generation}

When the Scala compiler ensures that the rules composition has valid types, it
expands the \emphcode{rule} macros to the code that would be run at runtime. Next,
we will describe all the steps from the rule definition to its code generation.

Consider this rule:

\begin{lstlisting}[language=Scala, showstringspaces=false]
val arule = rule { "ab" }
\end{lstlisting}

\emphcode{rule} on the right of the equal sign is the Scala method defined as:

\begin{lstlisting}[language=Scala, showstringspaces=false]
def rule[I <: HList, O <: HList](r: Rule[I, O]): Rule[I, O] =
  macro ParserMacros.ruleImpl[I, O]
\end{lstlisting}

\emphcode{"ab"} of the \emphcode{String} type is neither the instance nor the
subtype of type \emphcode{Rule[I,O]}. The Scala compiler succeeds in finding
the implicit: for this purpose, \emph{parboiled2} defines implicit conversion
from the \emphcode{String} type to the \emphcode{Rule0} type in \emphcode{Parser} class:

\begin{lstlisting}[language=Scala, showstringspaces=false]
@compileTimeOnly("Calls to `str` must be inside `rule` macro")
implicit def str(s: String): Rule0 = `n/a`
\end{lstlisting}

After being applied, the implicit turns \emphcode{arule} to:

\begin{lstlisting}[language=Scala, showstringspaces=false]
val arule: Rule0 = rule { SimpleParser.this.str("ab"): Rule0 }
\end{lstlisting}

Both \emphcode{@compileTimeOnly} and \emphcode{`n/a`} (a method that throws
\emphcode{IllegalStateException}) of the \emphcode{rule} method guards the
runtime execution from leaking of the \emphcode{str} method: if the execution path
somehow reaches the \emphcode{str} at runtime, it throws an exception. The macro
definition of the \emphcode{rule} method evaporates all the definitions of
the \emphcode{str} method.

The \emphcode{rule} method calls \emphcode{ParserMacros.ruleImpl[I,O]}.
\emphcode{ruleImpl} is a special kind of the method: it takes the Scala AST as an input,
transforms it, and returns the transformed Scala AST:

\begin{lstlisting}[language=Scala, showstringspaces=false]
def ruleImpl[I <: HList: ctx.WeakTypeTag,
             O <: HList: ctx.WeakTypeTag]
  (ctx:ParserContext)
  (r: ctx.Expr[Rule[I, O]]): ctx.Expr[Rule[I, O]] = {
    // ...
    val opTreeCtx = new OpTreeContext[ctx.type] {
      val c: ctx.type = ctx
    }
    opTreeCtx.OpTree(r.tree)
    // ...
  }
\end{lstlisting}

The dedicated trait \emphcode{OpTree} handles all possible rules
transformations. \emphcode{ruleImpl} creates its instance and passes the AST
\emphcode{r.tree} to it. The \emphcode{opTreePF} method transforms a definition in
the grammar to actual code:

\begin{lstlisting}[language=Scala, showstringspaces=false]
val opTreePF: PartialFunction[Tree, Tree] = {
  // ...
  case q"$a.this.str($s)" =>
    q"""
      val matched =
        input.sliceString(cursor, cursor + $s.length) == $s
      if (matched) cursor += $s.length
      matched
    """
  // ...
}
\end{lstlisting}

When the case pattern is applied to the expression
\emphcode{SimpleParser.this.str("ab")}, the values of \emphcode{a} and \emphcode{s} on the
right hand side are respectively \emphcode{SimpleParser} and \emphcode{"ab"}.
The naive implementation should do three things:

\begin{enumerate}
  \item compare the input slice to the string \emphcode{"ab"}

  \item if the input matches, it advances the \emphcode{cursor}
    (Section~\ref{rules-dsl}) further to the length of \emphcode{"ab"}

  \item return the \emphcode{Boolean} result of the match
\end{enumerate}

\emphcode{opTreePF} matches not only primitives, but complex rules operands as
well. Consider the \emphcode{firstOf} rule that is naturally coded as follows:

\begin{lstlisting}[language=Scala, showstringspaces=false]
val opTreePF: PartialFunction[Tree, Tree] = {
  case q"$lhs.|[$a, $b]($rhs)" =>
    q"""
      val cursorCurrent = cursor
      if (lhs()) true
      else {
        cursor = cursorCurrent
        if (rhs()) true
        else {
          cursor = cursorCurrent
          false
        }
      }
    """
}
\end{lstlisting}

\emphcode{lhs} and \emphcode{rhs} are the rules that could be composed of
primitives, other combinators, and other rule calls. In the end, they are
callable and return \emphcode{Boolean}. For example, in case of

\begin{lstlisting}[language=Scala, showstringspaces=false]
("a" ~ "b") | arule
\end{lstlisting}

the values of \emphcode{lhs} and \emphcode{rhs} would be
\lstinline[language=Scala]{("a" ~ "b")} and \lstinline[language=Scala]{arule}
respectively.

\subsection{Optimizations}

The naive implementation generates a string slice on every match attempt. A
possible optimization towards the efficient implementation would be
a char-by-char comparison in the imperative style:

\begin{lstlisting}[language=Scala, showstringspaces=false]
case q"$a.this.str($s)" => q"""
  var ix = 0
  while (ix < $s.length && cursorChar == $s.charAt(ix)) {
    ix += 1
    cursor += 1
  }
  ix == $s.length"""
\end{lstlisting}

The next optimization step comes from the observation that in most cases a
grammar contains domain specific string literals known at compile time. A
string literal, e.g. \emphcode{"abc"}, is unrolled to the nested list of
if/else-s as follows:

\begin{lstlisting}[language=Scala, showstringspaces=false]
if (cursorChar == 'a') {
    cursor += 1
    if (cursorChar == 'b') {
        cursor += 1
        cursorChar == 'c'
    } else false
} else false
\end{lstlisting}

Generally, \emphcode{opTreePF} in the first turn checks if \emphcode{s} is
a literal string and then applies the \emphcode{unroll} function that generates
if/else-s cascade:

\begin{lstlisting}[language=Scala, showstringspaces=false]
def unroll(s: String, ix: Int = 0): Tree =
  if (ix < s.length) q"""
    if (cursorChar == ${s charAt ix}) {
      cursor += 1
      ${unroll(s, ix + 1)}
    } else false
  """ else q"true"

case q"$a.this.str($s)" => s match {
  case Literal(Constant(sc: String)) => unroll(sc)
  case _ => // imperative code for general string match
\end{lstlisting}

Note how \emphcode{unroll} mixes the code generation logic with assertions of
what \emphcode{s} is at compile-time.

\emph{parboiled2} applies a few more optimizations as follows:
\begin{itemize}
  \item flatten a tree of \emphcode{sequence} rules series

  \item same technique for \emphcode{firstOf} rules series

  \item character sets (\emphcode{CharPredicate}). They allow to
    determine if in the input character belongs to the set. \emph{parboiled2} comes
    with plenty of predefined sets (like \emphcode{CharPredicate.Digit} and
    \emphcode{CharPredicate.Alpha}), and allows defining it from a function
    of the \emphcode{Char -> Boolean} type

\end{itemize}

\subsection{Code Generation Limitation} \label{code-generation-limitation}

The general limitation in a wider spread of effective code generation and
optimizations lies in the nature of Scala macros: the \emphcode{rule} macro
can only analyze the scope of a single method. Consider the grammar:

\begin{lstlisting}[language=Scala, showstringspaces=false]
val arule  = rule { "a" }
val aarule = rule { arule ~ arule }
\end{lstlisting}

Theoretically obvious optimization of \emphcode{aarule} is to inline
\emphcode{arule} and squash sequence of two \emphcode{"a"}s to the single
string \emphcode{"aa"}. But actually \emphcode{opTreePF} only sees the
\emphcode{arule} call without any non-hackable way to get the AST of
the \emphcode{arule} body.

\section{Catching Parsing Errors} \label{catching-parsing-errors}

An important part of the parsing process is error reporting: to identify
why the parsing failed and at what position. The Scala compiler generates a code only
once during the compilation. The exact same code should parse the input and
inform whether it fails to parse and why. The process distinguishes two major
phases:

\begin{itemize}

  \item the parsing phase:

\begin{lstlisting}[language=Scala, showstringspaces=false]
if (phase0_initialRun())
  scheme.success(valueStack.toHList[L]())
\end{lstlisting}

    If it successfully finishes, \emphcode{run} returns the top value on the
    \emphcode{ValueStack} wrapped in a successful result of the delivery scheme
    (Section~\ref{delivery_scheme-parsing_result})

  \item if it fails, next phases upon run determine the error index of the
    input and collect the rule traces. Each phase respects the rules that are marked as
    quiet. Finally, \emphcode{parserError} is wrapped in the error result of
    the delivery scheme:

\begin{lstlisting}[language=Scala, showstringspaces=false]
else {
  val principalErrorIndex: Int =
    phase1_establishPrincipalErrorIndex()
  val parseError: ParseError = // rest phases
  scheme.parseError(parseError)
}
\end{lstlisting}
\end{itemize}

A rule should return the tracing information when the execution path reaches it.
There is no code yet that preserves the tracing information
(Section~\ref{code-generation}). A rule code generation is encapsulated in a
reciprocal class. The class has two versions of code rendering: for the parsing
phase and for the error collecting phase. Consider the \emphcode{CharMatch} class
for the basic char rule:

\begin{lstlisting}[language=Scala, showstringspaces=false]
case class CharMatch(charTree: Tree) extends TerminalOpTree {
  def ruleTraceTerminal =
    q"org.parboiled2.RuleTrace.CharMatch($charTree)"
  def renderInner(wrapped: Boolean): Tree = {
    val unwrappedTree = q"cursorChar == $charTree && __advance()"
    if (wrapped)
      q"$unwrappedTree && __updateMaxCursor() || __registerMismatch()"
    else unwrappedTree
  }
}
\end{lstlisting}

The \emphcode{unwrappedTree} has a code described in
Section~\ref{code-generation}. The addendum is that \emphcode{CharMatch}
renders based on the \emphcode{wrapped} flag. And the wrapped version should either
update the max cursor if it matches, or register a mismatch.
\emphcode{TerminalOpTree} implements the mismatch registration and the error tracing
information.

\section{Further Work} \label{further-work}

Obstacles to wider optimizations originate in the narrow scope of the rule macro
application, as mentioned in Section~\ref{code-generation-limitation}.
Notably, it blocks cross-rule optimizations and indirectly increases the code
base. For example, \cite{Ford:2004:PEG:982962.964011} theoretically
showed that \emphcode{oneOrMore}, \emphcode{option}, and
\emphcode{and-predicate} operators are ``syntactic sugar'', i.e. the combination
of other operators that can substitute them. Staging and compilation
techniques~\cite{rompf2010lightweight} might evaporate intermediate data
creation. But they require a much wider scope. And \emph{parboiled2} should
explicitly implement ``syntactic sugar'' operators individually for the sake of
runtime effectiveness.

Single method code generation by macro also limits a code block to handle all
the facilities (like debugging and tracing). Such code generation potentially blow up
the method code size (limited by JVM), complicate the code base support, and lessen
\emph{parboiled2} versions back compatibility.

Another problem arises from the fact that the \emphcode{rule} macro depends on
the context that it does not control. For example, a rule might be assigned either to
\emphcode{val} or \emphcode{def}. Both approaches have pros and cons. But we
should make design decisions that define inner implementation and library
usages. This is another point where backward compatibility suffers.

Creating higher-ordered rules (a method that takes another rule as a parameter)
is also impossible with the current version of Scala Macros.

The origins of the \emphcode{ValueStack} arise from the inefficiency of the combinator
approaches -- they produce too many intermediate data structures. The first
negative thing is \emph{parboiled2} shifts side-effect result composition too much on
the developers' shoulders. Hence, again we are constrained with the API and
backward compatibility. Another drawback is the \emphcode{ValueStack}
type-based verification, which is good for the user when the type check passes. If a user
makes a mistake somewhere in typing (i.e., missed an argument type in lambda for
action operation), the Scala compiler fires tens of lines of machine-generated
typing errors that are really hard to interpret by a human.
\cite{Jonnalagedda2014} showed how to eliminate
intermediate data structures automatically.

The described limitations restrict intuitive feature implementation: creating
custom rules that need inner API access. For example, it is hard to implement a
rule that tracks position coordinates of parsed AST nodes.

The good news is that a new version of Scala Macros should be sufficient to overcome
all the obstacles \cite{liu2017two}.

\section{Acknowledgments}

The core development is supported by Google Summer of Code 2013 grant.
The development process leading and significant contributions was made
by Mathias Doenitz.

The work on the paper is supported by the National Science Foundation (NSF~DBI-1356347).

\bibliography{parboiled2}

% \todototoc
% \listoftodos

\end{document}